\documentclass[aps,prb,twocolumn,superscriptaddress]{revtex4-2}
\usepackage{amssymb, amsmath, bm}
\usepackage{graphicx,onlyamsmath}
\usepackage[utf8]{inputenc}
\usepackage{siunitx}
\usepackage[version=4]{mhchem}

\usepackage{natbib}
\usepackage{xcolor}
\usepackage[normalem]{ulem}  % ��� ����������� ������
\UseRawInputEncoding

\usepackage{mathtools}
\usepackage{graphicx}
\usepackage{mathrsfs}
\usepackage{amssymb}
\usepackage{xcolor}
\usepackage{amsmath}
\usepackage{bm}
\usepackage{amsmath}
\usepackage{physics}
\makeatletter\AtBeginDocument{\let\@elt\relax}\makeatother
\usepackage{amsmath}
\usepackage[normalem]{ulem}
\usepackage[unicode=true,colorlinks=true,citecolor=blue,urlcolor=blue]{hyperref}
\usepackage{makecell}

\def\be {\begin{equation}}
	\def\ee {\end{equation}}
\def\bea {\begin{align}}
	\def\eea {\end{align}}

\def\bee{\begin{eqnarray}}
	\def\eee{\end{eqnarray}}
\def\BC {\begin{cases}}
	\def\EC {\end{cases}}

\makeatother

\begin{document}

\title{Strongly nonlinear Bernstein modes in graphene reveal plasmon-enhanced near-field magnetoabsorption \\
%	\re{Pressure-Driven  Positive Photoconductivity in  CdHgTe Structures}
}

\author{I.~Yahniuk}
\affiliation{Institute of Experimental and Applied Physics, University of Regensburg, 93040 	Regensburg, Germany}
\affiliation{Institute of High Pressure Physics of the Polish Academy of Sciences, 01-142 Warsaw, Poland}

\author{I.A.~Dmitriev}
\affiliation{Institute of Experimental and Applied Physics, University of Regensburg, 93040 	Regensburg, Germany}

\author{A.L.~Shilov}
\affiliation{Department of Materials Science and Engineering, National University of Singapore, 117575 Singapore}

\author{E.~M{\"o}nch}
\affiliation{Institute of Experimental and Applied Physics, University of Regensburg, 93040 Regensburg, Germany}

\author{M.~Marocko}
\affiliation{Institute of Experimental and Applied Physics, University of Regensburg, 93040 Regensburg, Germany}

\author{J.~Eroms}
\affiliation{Institute of Experimental and Applied Physics, University of Regensburg, 93040 Regensburg, Germany}

\author{D.~Weiss}
\affiliation{Institute of Experimental and Applied Physics, University of Regensburg, 93040 Regensburg, Germany}

\author{P.~Sadovyi}
\affiliation{Institute of High Pressure Physics of the Polish Academy of Sciences, 01-142 Warsaw, Poland}

\author{B.~Sadovyi}
\affiliation{Institute of High Pressure Physics of the Polish Academy of Sciences, 01-142 Warsaw, Poland}
\affiliation{Faculty of Physics, Ivan Franko National University of Lviv, 79005, Lviv, Ukraine}

\author{I.~Grzegory}
\affiliation{Institute of High Pressure Physics of the Polish Academy of Sciences, 01-142 Warsaw, Poland}

\author{W.~Knap}
\affiliation{Institute of High Pressure Physics of the Polish Academy of Sciences, 01-142 Warsaw, Poland}
\affiliation{CENTERA, CEZAMAT, Warsaw University of Technology, 02-822 Warsaw, Poland}

\author{J.~Gumenjuk-Sichevska}
\affiliation{Johannes Gutenberg-University Mainz, D-55128 Mainz, Germany}
\affiliation{V. Lashkaryov Institute of Semiconductor Physics, National Academy of Science, 03028, Kyiv, Ukraine}

\author{J.~Wunderlich}
\affiliation{Institute of Experimental and Applied Physics, University of Regensburg, 93040
	Regensburg, Germany}
%\alsoaffiliation{Institute of Physics, Czech Academy of Sciences, Cukrovarnická 10, 162 00 Praha 6, Czech Republic}

\author{D.~A.~Bandurin}
\affiliation{Department of Materials Science and Engineering, National University of Singapore, 117575 Singapore}

\author{S.~D.~Ganichev}
\affiliation{Institute of Experimental and Applied Physics, University of Regensburg, 93040 	Regensburg, Germany}
\affiliation{Institute of High Pressure Physics of the Polish Academy of Sciences, 01-142 Warsaw, Poland}

%	Warsaw:
%	
%	1,2 Yurii Ivonyak (Yurii.Ivonyak@unipress.waw.pl) –  all work related to pressure cells was performed in collaboration with him.
%	
%	1 Bercha Artem (artem@unipress.waw.pl) – I and prof.  Witold Trzeciakowski agreed that I can use the pressure cells and equipment from his laboratory, however, I should add ’’Bercha Artem’’ to the list of co-authors. !!! all pressure work was done in his lab.
%	
%	1,2  Cywiński Grzegorz  (grzegorz.cywinski@unipress.waw.pl) - I would add him due to the formal way, because he always helps with documents and formal issues.
%	
%	1 Institute of High Pressure Physics, Polish Academy of Sciences, ul. Sokołowska 29/37, 01-142 Warsaw, Poland
%	
%	2 CENTERA, CEZAMAT, Warsaw University of Technology, ul. Poleczki 19, 02-822 Warsaw, Poland
%	
%	

%\date{\today}% It is always \today, today,
       %  but any date may be explicitly specified

\begin{abstract}
Bernstein modes --- hybrid magnetoplasmon excitations arising from the coupling between cyclotron motion and collective oscillations in two-dimensional electron systems—offer direct access to non-local electrodynamics. These modes can exhibit rich nonlinear behavior akin to strong-coupling phenomena in cavity quantum electrodynamics, but reaching nonlinear regime has remained experimentally challenging.  Here we report the observation of nonlinear Bernstein modes in graphene using terahertz excitation with near-field enhancement from embedded metallic contacts. Photoresistance spectroscopy reveals sharp resonances at $B_c/2$ and $B_c/3$ that saturate at radiation intensities nearly an order of magnitude lower than the cyclotron resonance. We ascribe this to strong local heating of the electron gas due to resonant excitation of high-amplitude Bernstein magnetoplasmons, associated with a combination of the field-concentration effect of the near field and plasmonic amplification that is resonantly enhanced in the region of Bernstein gaps. Polarization-resolved measurements further confirm the near-field origin: Bernstein resonances are insensitive to circular helicity but strongly depend on the angle of linear polarization, in sharp contrast to the cyclotron resonance response. Our results establish  graphene as a platform for nonlinear magnetoplasmonics, opening opportunities for strong-field manipulation of collective electron dynamics, out-of-equilibrium electron transport, and solid-state analogues of cavity quantum electrodynamics.
\end{abstract}

 % PACS, the Physics and Astronomy
                             % Classification Scheme.
%Use showkeys class option if keyword                            %display desired
\maketitle

\section{\label{sec:Int}Introduction}

\begin{figure*}[t]
	\centering
	\includegraphics[width=\linewidth]{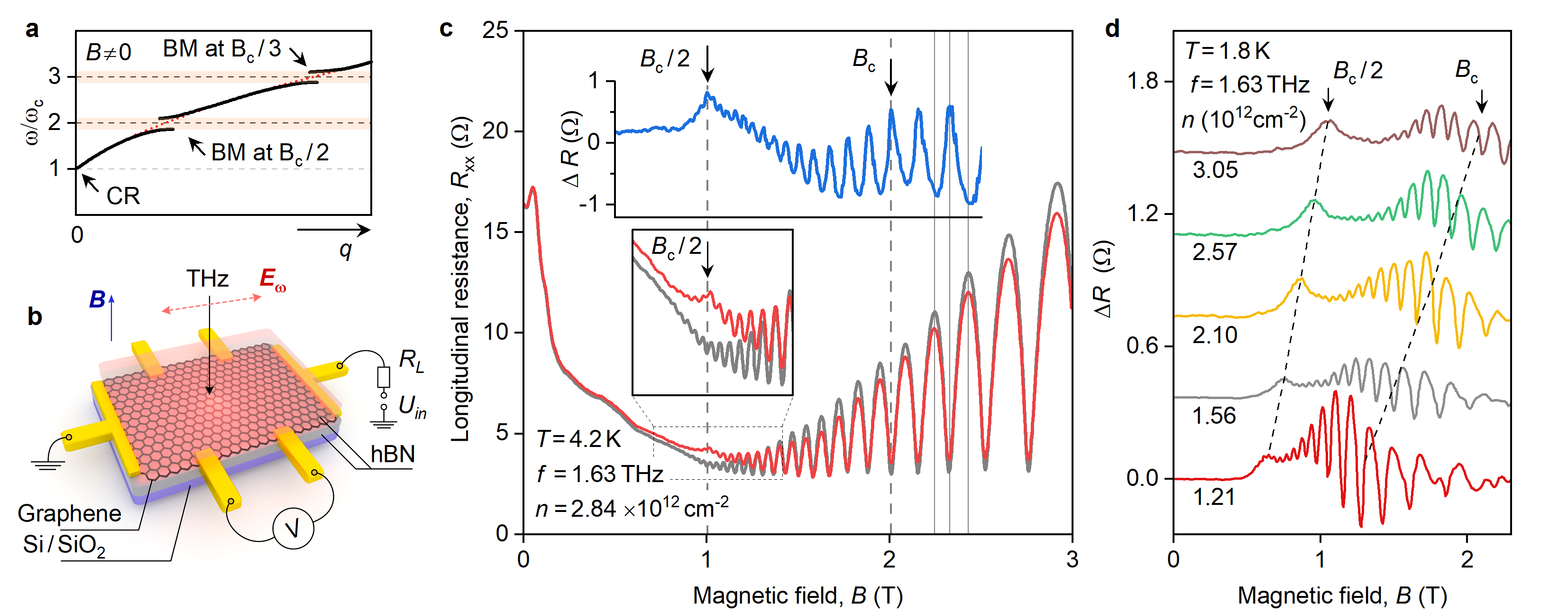}
	\caption{ (a) Magnetoplasmon dispersion (schematic): due to non-local effects at $1/q$ comparable to the spatial scale of the cyclotron motion, it includes the avoided crossings and spectral gaps at $\omega/\omega_c\simeq N$, $N=2,3,..$, corresponding to $B=B_c/N$, in contrast to the conventional magnetoplasmon in the local approximation, red dashed line, following $\omega^2=\omega_c^2+\omega_p^2$ where $\omega_p$ is the plasma frequency at $B=0$. The arrow at $q=0$ indicates the collective uniform CR mode at $\omega=\omega_c$ ($B=B_c$). (b) Measurement setup: The longitudinal resistance $R_{xx}$ %\re{(4-terminal; measured on edge contacts)} 
		is measured in a single-layer graphene sample with embedded metallic contacts
		%\re{(4-terminal, see panel b)}, 
		subjected to a magnetic field $B$ and normally incident {\it cw} THz radiation (Faraday configuration). (c) Longitudinal resistance $R_{xx}$  
		%\re{(2-terminal, source-drain)}
		% \re{(2-terminal; measured on source-drain contacts)}
		as a function of $B$ in the presence (red line) and absence (gray line) of THz irradiation at a frequency of $f =\omega/2\pi= 1.63$~THz, measured at $T=4.2$~K and back-gate voltage corresponding to an electron carrier concentration $n= 2.84\times10^{12}$~cm$^{-2}$. The lower inset shows a zoomed-in view of the peak at $B_{c}/2$. The upper inset represents the photoresistance $\Delta R$ -- the difference in  $R_\text{xx}$ in the presence and absence of the THz radiation. The vertical dashed gray lines indicate the positions of the CR, $B_c$, and its main overtone, $B_c/2$.  Thin vertical lines illustrate the opposite phase of oscillations in $\Delta R$ with respect to $R_\text{xx}$, corresponding to suppression of the SdH oscillations under THz illumination. (d) Photoresistance $\Delta R$ versus the magnetic field, measured at $T=1.8$ K and $f=1.63$ THz for various carrier densities $n$, extracted from the period of SdH oscillations. The dashed lines indicate the positions of the CR and its main overtone at the corresponding densities. The intensity of the incoming radiation was $I=0.165$~W/cm$^2$.
	}
	\label{fig1}
\end{figure*}

The electromagnetic response of two-dimensional electrons in magnetic fields is traditionally described by cyclotron resonance (CR), in which electrons execute circular orbits at the cyclotron frequency $\omega_c$ close to the excitation frequency $\omega$. This picture changes qualitatively at the nanoscale, when the excitation frequency matches that of plasmons. The coupling of plasmons with cyclotron motion reshapes the magnetoplasmon spectrum and splits the otherwise continuous dispersion into a series of hybrid excitations known as Bernstein modes (BMs)~\cite{Bernstein_1958}. While plasmons and BMs were extensively studied and are well understood in high-mobility III–V semiconductors~\cite{Sitenko1957,Bernstein_1958,Quinn_1974_Bernstein2DES,BMBatke,Gudmundsson_1995_Bernstein_modes_wires_dots,BMBatke2,BM_photoconductivity,Volkov2014,Dorozhkin2021,Yavorskiy2025}, their investigation has been revitalized by the advent of graphene and van der Waals heterostructures~\cite{ni2018fundamental,Koppens_2015_ConfinedLowLoss}. Plasmonic experiments in these systems have provided unique access to hidden properties of low-dimensional electrons and their collective excitations~\cite{Fogler}. Near-field probes have revealed a range of novel effects, including many-body interactions that reshape plasmon dispersion~\cite{Rostami-2017,Lundeberg2017}, electron–hole sound modes~\cite{hydro-plasmons1}, hydrodynamic plasmons~\cite{hydro-plasmons2}, and quasi-relativistic Fizeau drag~\cite{Fizeau1,Fizeau2}. Against this backdrop, Bernstein modes in graphene~\cite{Bernstein-modes-graphene,BandurinBernstein} provide a distinct perspective: their temperature dependence revealed signatures of tomographic electron hydrodynamics~\cite{tomographic2}, signaling a striking departure from conventional Fermi-liquid behavior in two dimensions~\cite{Ledwith}. Moreover, owing to avoided crossings in their dispersion (Fig.~\ref{fig1}a), BMs resemble the strong coupling regime of hybrid light–matter systems familiar from cavity quantum electrodynamics~\cite{Scalari_Ultrastrong_CR_THZ,Muravev_StrongCoupling}. Such hybridization opens the door to nonlinear phenomena, potentially giving rise to effects such as Rabi oscillations~\cite{Belyanin_Time-resolved_CR}, electromagnetically induced transparency, or out-of-equilibrium manifestations of tomographic hydrodynamics~\cite{Ledwith,Tomography_Levitov2}. Realizing this, however, requires access to the nonlinear regime, which has not yet been achieved.

In this work, we demonstrate strong nonlinearity of BMs in graphene under continuous-wave terahertz (THz) excitation combined with BM-assisted resonant near-field enhancement. The local THz field, concentrated near metallic contacts penetrating the graphene sample area (Fig.~\ref{fig1}b), is strongly modified and resonantly enhanced when the THz frequency approaches the regions of nearly flat dispersion around the BM gaps in the magnetoplasmon dispersion (Fig.~\ref{fig1}a).\cite{Bernstein-modes-graphene} The enhanced local field, akin to the lightning-rod effect, pushes the system well beyond the linear regime. We compare the nonlinear saturation of the BM signal to the simultaneously observed CR, coming from the uniform part of the THz wave illuminating the sample, and find that the strong nonlinearity is specific to the near field-generated BMs. The near-field character of the BM resonances is additionally evidenced by their specific responses to the circularly and linearly polarized THz radiation, which are qualitatively distinct from those of the CR. Our findings open a new direction in exploring nonlinear, non-local electrodynamics in 2D materials.

\section{Experimental results}
\label{Experimental results}

%{\bf Experimental results.}
The Bernstein resonances are observed in monolayer graphene excited by continuous wave (cw) THz laser radiation. The graphene sample was prepared  using the technology described in Ref.~\cite{Sadovyi2025} and in Supplementary Materials. The coupling of the incoming plane THz wave to the short-wavelength Bernstein modes is mediated by metallic contacts embedded into the sample area;\cite{BandurinBernstein,tomographic2} see Fig.~\ref{fig1}b and Methods. Figure~\ref{fig1}c presents a typical magnetic field ($B$) dependence of the low-temperature ($T$) longitudinal resistance in the presence (red) and absence (gray) of the THz irradiation. The most appealing effect of illumination is the appearance of a sharp peak at $B=1$ T, highlighted in the inset. As expected for the principal BM at $\omega\simeq2\omega_c$ (Fig.~\ref{fig1}a), the position of this peak corresponds to half ($B_c/2$) of the CR position, $B_c=m_c \omega/e$, at this carrier density ($n$), see arrows in Fig.~\ref{fig1}c. Here, the cyclotron mass, $m_c= \hbar \sqrt{\pi n}/v_F$, describes the spacing $\hbar\omega_c=\hbar e B/m_c$ between Landau levels (LLs) at the Fermi energy $E_F$ in the semiclassical regime of high Landau levels $E_F\gg \hbar\omega, \hbar\omega_c, k_B T$, relevant for our experiments, and $v_F=10^6$~m/s is the Fermi velocity in graphene. Apart from the BM peak, at higher magnetic fields, the irradiation leads to the conventional suppression of the Shubnikov-de Haas (SdH) oscillations due to electron gas heating, which is maximized in the vicinity of the CR at $B\sim B_c$. 

In the following, we focus on the photoresistance, $\Delta R$, defined as the change in $R_\text{xx}$ under THz illumination (blue line in Fig.~\ref{fig1}c). In the photoresistance, BM excitation produces a sharp peak at $B_c/2$, while the suppression of the SdH oscillations near the CR gives rise to magnetooscillations with the phase opposite to that in $R_\text{xx}$ (dashed lines in Fig.~\ref{fig1}c). The CR response is further shaped by the exponential growth of SdH oscillations, which obscures its purely resonant character. The evolution of $\Delta R$ with the carrier density is shown in Fig.~\ref{fig1}d. Here and below the photoresistance is measured applying double-modulation technique~\cite{Kozlov2011,Otteneder2018,Savchenko2021} and extracted from the photosignal $V$ as $\Delta R=V R_L/\mid U_\text{in} \mid$, where $U_\text{in}$ is a bias voltage (see Sec. Methods). The CR signal is strongest at low densities, whereas the BM amplitude shows only a weak dependence on density. The positions of both resonances follow the density scaling $B_c\propto\sqrt{n}$ expected for the linear dispersion of graphene in the semiclassical regime. This is illustrated by dashed lines showing the calculated positions of $B_c/2$ and $B_c$, obtained using the electron density $n$ extracted from the period of the SdH oscillations.

\begin{figure*}[t] 
	\centering
	\includegraphics[width=\linewidth]{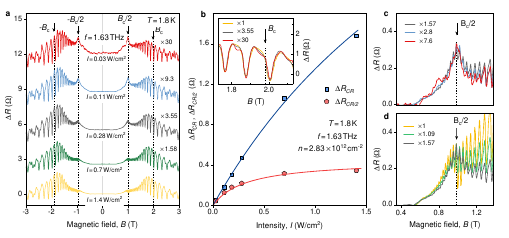}
	\caption{  (a) Photoresistance $\Delta R$ vs. magnetic field $B$ at varied intensities $I$ of incoming $f=1.63$ THz irradiation, as given under each trace, measured at $T=1.8$~K and fixed $n=2.83\times{10^{12}}$~cm$^{-2}$. Dashed lines and arrows show the positions of the CR, $\pm B_c$, and its 2nd harmonics, $\pm B_c$/2. Individual traces are shifted and scaled for clarity. The scaling factors (on the right of each trace) are chosen such that the amplitude of magnetooscillations near the CR, $\Delta R_\text{CR}$, remains the same. This is illustrated in the inset in (b); the color code here and in (c), (d) corresponds to that in (a). The resulting amplitudes $\Delta R_\text{CR}$ at corresponding $I$ are shown using blue square symbols in (b). As evident from (a), the BM photoresponse near $\pm B_c/2$ features a quite different evolution with intensity $I$. (c) and (d) show three lowest-$I$ and three highest-$I$ curves from (a), respectively; the gray trace is shown in both (c) and (d). Here, the scaling factors are chosen such that the height of the BM resonance, $\Delta R_\text{CR/2}$, remains the same. The magnitudes of the BM signal $\Delta R_\text{CR/2}$ at corresponding $I$ are shown using red pentagon symbols in (b) and demonstrate almost full saturation of the BM photoresponse. Solid lines in (b) are fits using Eq.~\eqref{formula}.
		}  
	\label{fig2}
\end{figure*}

Having introduced the typical appearance of BMs in our experiments (see also \cite{BandurinBernstein,tomographic2}), we turn to our central observation: a striking nonlinearity of the low-$T$ BM-assisted photoresistance as a function of the intensity $I$ of the incoming THz radiation. Figure~\ref{fig2}a presents  photoresistance dependence on $B$ measured at $T = 1.8$~K under 1.63 THz radiation for representative intensities spanning almost two orders of magnitude. The individual traces are vertically shifted and rescaled to make the evolution of the photoresponse clearly visible. The multiplication factors shown to the right of the traces are chosen such that the magnitude of the photoresponse in the vicinity of the CR, $B\sim \pm B_c$, is the same. In such a presentation the magnitude of the Bernstein peaks at $B\sim \pm B_c/2$ is drastically reduced at high radiation intensities. Even without further analysis, the distinct behaviors of the cyclotron and Bernstein resonances reveal the high nonlinearity of the photoresponse. 
Indeed, the intensity dependence of the magnitude of the Bernstein photoresponse, $\Delta R_\text{CR/2}$ at $B_c/2$ (shown by red symbols in Fig.~\ref{fig2}b) demonstrates nearly complete saturation. Although such nonlinear behavior could be obtained via measurement of the intensity dependence of $\Delta R$ at $B_c/2$, we performed a more detailed analysis that also traces the changes in the shape of the Bernstein resonance with intensity in the presence of quantum oscillations. To this end, we have taken the three lowest-intensity traces from Fig.~\ref{fig2}a and found that they can be well rescaled to each other via a proper choice of constant multiplication factors, see Fig.~\ref{fig2}c. The shape of the Bernstein resonance at the low-$B$ tail does not change with intensity, whereas the high-$B$ tail is progressively stronger affected by SdH oscillations as intensity grows. Further increase of $I$ makes the relative contribution of oscillations even stronger, as demonstrated in Fig.~\ref{fig2}d. Note that the shape of the low-$B$ tail of the resonance still remains unchanged. The multiplication factors in Figs.~\ref{fig2}c and~\ref{fig2}d determine the positions of the red symbols in Fig.~\ref{fig2}b. By contrast, the photoresponse at the CR scales almost linearly with intensity, see blue symbols in Fig.~\ref{fig2}b. Consistently, the shape of oscillations in $\Delta R$ near the CR does not change with intensity, as illustrated by three rescaled traces in the inset of Fig.~\ref{fig2}b.

\begin{figure*}[t] 
	\centering
	\includegraphics[width=\linewidth]{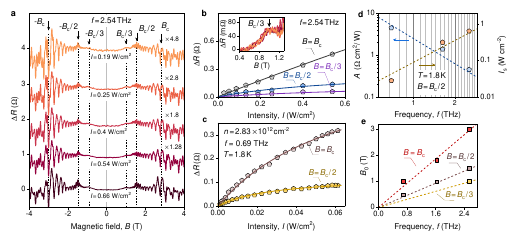}
	\caption{  (a) Photoresistance $\Delta R$ vs. magnetic field $B$ at varied intensities $I$ of incoming $f=2.54$ THz irradiation. Other parameters and notations are the same as in Fig.~\ref{fig2}. In addition to the primary BM resonance at $B_c/2$, at higher frequency also the next BM resonance at $B_c/3$ can be resolved. The rescaled traces near $B_c/3$ are shown in the inset in (b). The corresponding intensity dependence of the photoresponse at $B_c/3$ is represented by purple pentagons in (b); blue and gray pentagons show the  intensity evolution of the resonances at $B_c/2$ and $B_c$. Zoomed-in plots demonstrating saturation of the signals for $B_c/2$ and $B_c/3$ are presented in the Supplemental Materials. (c) Intensity dependence of the photoresistance at the CR, $B_c$, and at the primary BM resonance, $B_c/2$, for a lower frequency of $f=0.69$ THz. Solid lines in (b) and (c) are fits using Eq.~\eqref{formula}. The corresponding fitting parameters $A(f)$ and $I_s(f)$ are presented in (d) using a double-logarithmic scale; here we also include the results $f=1.63$ THz from Fig.~\ref{fig2}. Dashed lines following the $1/f^2$ behavior for $A(f)$ and $f^2$ for $I_s(f)$ are shown as guide for the eye. (e) Positions of the detected CR and BM resonances for all three frequencies are shown as symbols; dashed lines are the calculated positions of $B_c$, $B_c/2$, and $B_c/3$ using the density $n=2.83\times{10^{12}}$ cm$^{-2}$ determined from the period of the SdH oscillations.
		}
	\label{fig3}
\end{figure*}

The nonlinearities of Bernstein resonances were studied for several THz frequencies. The data for a higher frequency, $f= $2.54 THz, are shown in Fig.~\ref{fig3}a. Here, apart from the principal BM resonance at $B_c/2$, the next BM at $B_c/3$ becomes clearly detectable, see also inset of Fig.~\ref{fig3}b. At this frequency and within the available range of intensities, the nonlinearity of both Bernstein resonances is present but not pronounced (see Supplemental Materials), while the photoresponse near the CR remains fairly linear. In sharp contrast, at lower frequency, $f= $0.69 THz, the saturation of both BM and CR photoresponses is very pronounced, see Fig.~\ref{fig3}c. This is remarkable given that the available intensities at $f=0.69$~THz are more than an order of magnitude smaller than those at $f=0.69$~THz, see Figs.~\ref{fig3}c and~\ref{fig2}b. On the other hand, Fig.~\ref{fig3}e shows that the magnetic-field positions of all detected resonances scale linearly with the excitation frequency, in agreement with expectations for the semiclassical regime of our experiments. They follow the dashed lines in Fig.~\ref{fig3}e showing positions of $B_c=\hbar\omega \sqrt{\pi n}/e v_F$, $B_c/2$, and $B_c/3$ calculated using the electron density $n$ determined from the period of SdH oscillations.

We find that the observed nonlinearities at all frequencies can be well fitted by a phenomenological equation,\cite{Allan1985,Ganichev2005,Candussio2021a,Moldavskaya2025}
\begin{equation}
	\Delta R= \dfrac{A(f)I}{1+I/I_\text{s}(f)},
	\label{formula} 
\end{equation}
where the frequency-dependent parameters $A(f)$ and $I_\text{s}(f)$ characterize the photoresponse in the linear regime and the saturation intensity, respectively. The corresponding fits to the experimental data are shown as solid lines in Figs.~\ref{fig2}b and~\ref{fig3}b,c. Figure~\ref{fig3}d presents the frequency dependence of the extracted fitting parameters for the principal Bernstein resonance at $B=B_c/2$ using a double-logarithmic scale. It reveals that the coefficient $A(f)$ of the linear response decreases with $f$ (blue symbols), whereas the saturation intensity $I_s(f)$ grows with $f$ (orange symbols). Although the results are limited to three frequencies, it is evident that both parameters have a strong frequency dependence. Based on these results, the dependence appears roughly quadratic for $I_s(f)$ and inverse quadratic for $A(f)$, see dashed lines in Fig.~\ref{fig3}d. Regardless of the exact functional dependencies, which can be sensitive to the details of near-field formation, such as contact shape, and require more systematic studies, an important finding is that the saturation intensity $I_s(f)$ behaves approximately as the inverse of the linear coefficient $A(f)$, see Discussion. 

Using such analysis, we can also quantify another important aspect of our findings, namely, we can comparatively study the typical intensities of the incoming THz wave causing the nonlinearity of the CR and BM photoresistance. Application 
of Eq.~\eqref{formula} to the photoresponse near the CR, see Figs.~\ref{fig2}b and~\ref{fig3}b,c, reveals that the saturation intensity $I_s(f)$ for the CR is always substantially larger than that for the BM resonance. Namely, the solid lines in Fig.~\ref{fig2}b  ($f=1.63$ THz) correspond to $I_s=3.0$~W/cm$^2$  at the CR, $\Delta R_\text{CR}$, which is 9.4 times larger than $I_s=0.32$~W/cm$^2$  for the BM resonance, $\Delta R_\text{CR/2}$. For $f=0.69$ THz in Fig.~\ref{fig3}c we get $I_s=0.066$~W/cm$^2$ for the CR and $I_s=0.028$~W/cm$^2$ for the BM resonance (ratio of 2.4). For the highest frequency, $f=2.54$ THz, the saturation intensity for the primary BM resonance at $B=B_c/2$ can be estimated as $I_s=0.65$~W/cm$^2$, similar to that for the BM at $B=B_c/3$, $I_s=1.15$W/cm$^2$, while in the region of the CR, the photoresistance remains fairly linear over the entire range of available $I$, see  Fig.~\ref{fig3}b, meaning the corresponding $I_s\sim 10$~W/cm$^2$ is again substantially larger than that for the BM resonances. These observations provide evidence for strong near-field enhancement associated with the resonant excitation of BMs, see Discussion.

\begin{figure*}[t]
	\centering
	\includegraphics[width=0.85\linewidth]{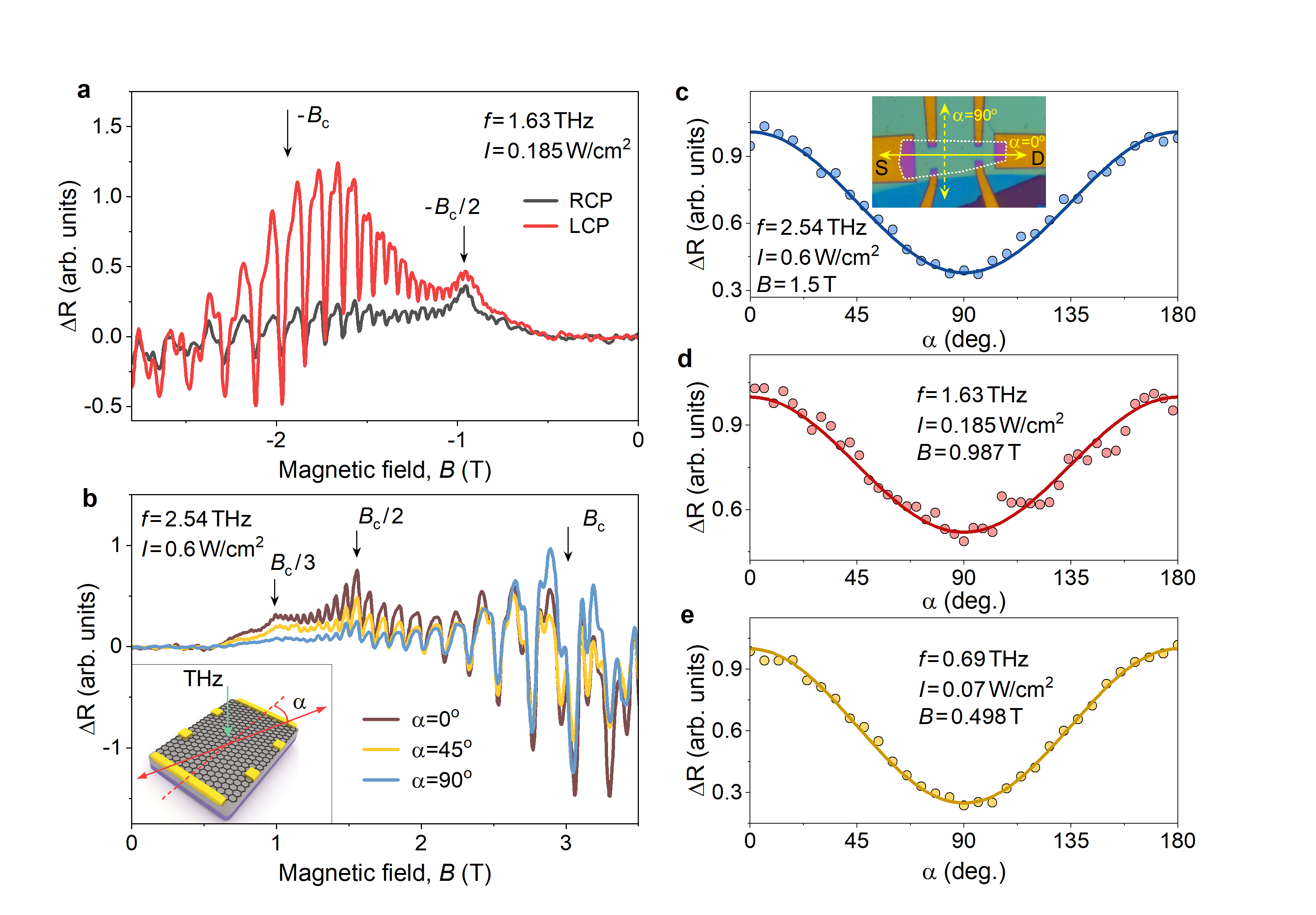}
	\caption{ %\re{(2-terminal; measured on source-drain contacts)}
		(a) Photoresistance $\Delta R$ vs. $B$ for left-circularly polarized (LCP, red) and right-circularly polarized (RCP, black) $f=1.63$ THz radiation. Other parameters and notations are the same as in Figs.~\ref{fig2} and \ref{fig3}. (b) Photoresistance $\Delta R$ vs. $B$ for different orientations of the electric field of linearly polarized $f=2.54$ THz radiation. The corresponding azimuthal angle of the THz field $\alpha$ is defined in the inset. (c)-(e) The $\alpha$-dependences of the BM photoresponse at $B=B_c/2$ for all three frequencies (symbols). The solid lines are fits using Eq.~\eqref{cos}. The inset in panel (c) shows a photograph of the studied structure. The white dashed line shows the region occupied by the graphene flake. The overlap of the contacts and graphene is marked in violet. 
			}
	\label{fig4}
\end{figure*}

Apart from the nonlinearity of the BM photoresponse, we have studied how it is affected by the polarization state of the incoming THz wave. Figure~\ref{fig4}a shows the photoresistance for right- and left-circularly polarized radiation. As expected, the CR photoresponse is well pronounced for the cyclotron-active radiation helicity only (red curve). In sharp contrast, the BM resonance has the same magnitude and shape for both helicities. A clearly distinct behavior of the CR and BM photosignals is also detected for the linearly polarized incoming radiation upon variation of the in-plane orientation of its electric field. Figure~\ref{fig4}b demonstrates that the photoresponse in the region of the CR is nearly unaffected by the orientation of the incoming radiation, characterized by the azimuth angle $\alpha$, defined in the inset. In contrast, the magnitude of the BM resonance turns out to be strongly $\alpha$-dependent. The measured dependences of the magnitude of $\Delta R$ at the peak corresponding to the principal BM resonance at $B=B_c/2$ are shown in Figs.~\ref{fig4}c-e for all three THz frequencies. All these dependencies can be well fitted by the equation
\begin{equation}
	\Delta R= a + b\cos^2{\alpha},  
	\label{cos}
\end{equation}
with two frequency-dependent fitting parameters $a$ and $b$.  
A photograph of the studied structure is presented in inset in Fig.~\ref{fig4}c. The data  were obtained using the source and drain contacts labeled S and D. The white dashed line shows the region occupied by the graphene flake. The overlap of the contacts and graphene is marked in violet. 
The observed distinct behavior of the CR and BM resonances can be directly attributed to the near-field origin of the BM resonances, see Discussion below.

%{\bf Discussion.} 

%Our observations simultaneously reveal CR and sharp resonances at the second and third harmonics of the CR. This is supported by the observed evolution of the peak positions with the electron density $n$ and frequency $f=\omega/2\pi$ of the THz radiation, see Figs.~1d and 3e. Indeed, in the semiclassical regime of high Landau levels (LLs), relevant for our experiments, the condition of the CR can be written as $B_c=\omega m_c/e$, where the cyclotron mass $m_c= \hbar \sqrt{\pi n}/v_F$ describes the distance $\hbar\omega_c=\hbar e B/m_c$ between LLs at the Fermi energy $E_F$ in the semiclassical regime $E_F\gg \hbar\omega, \hbar\omega_c, k_B T$ (here $v_F=10^6$~cm$^2$ is the Fermi velocity of graphene). Consequently, the non-equidistant spectrum of LLs in graphene reveals itself in the square-root density dependence of positions of CR and its harmonics, $B_c/N$, $N=1,2,3\ldots$, see Fig.~1d, while the frequency dependence in the semiclassical regime at fixed $n$ is linear, Fig.~3e. The calculated positions of resonances shown by dashed lines in these and other figures were obtained using the electron density $n$ extracted from the period of SdH oscillations.

\section{Discussion}
\label{discussion}
%{\bf Discussion.} 
Our observations simultaneously reveal the CR and sharp resonances at the second and third harmonics of the CR, attributed to the resonant near-field excitation of short-wavelength Bernstein magnetoplasmons. Our central finding is the observation of a strong nonlinearity of the BM photoresponse as a function of the intensity of the incoming THz radiation. The nonlinear regime has not been reached in earlier works on this subject. In the following discussion we first address the common physical mechanism of resonant absorption detection in the photoresistance experiments, related to electron heating, and then describe the nonlinearities, focusing on the BM-mediated near-field enhancement that causes fast saturation of the photoresistance at the BM resonances. 

The electron heating, or bolometric effect, is the main mechanism responsible for the detection of both CR and BM resonances in photoresistance within the semiclassical transport regime relevant to our experiments. In the case of CR, electron heating is associated with the scattering-assisted absorption of the homogeneous THz radiation, which is maximized at $B\sim B_c$.\cite{Seeger2004,Hilton2012} The uniform $q=0$ component of the in-plane THz electric field, produced by the homogeneous THz illumination of the sample at normal incidence, drives the uniform oscillatory motion of electrons throughout the sample.  The amplitude of this motion is maximized at the CR, $\omega=\omega_c$, corresponding to the long-wavelength limit $q\to0$ in the magnetoplasmon spectrum, see Fig.~\ref{fig1}a. 
%The amplitude of this motion is maximized at the CR, $\omega=\omega_c$, corresponding to the collective plasma modes at $q\to0$, see Fig.~\ref{fig1}a. 
In the presence of momentum relaxation, this leads to a resonant absorption of the uniform component of THz wave maximized at $B\sim B_c$. In the stationary state, the energy absorbed into the electron subsystem is balanced by its transfer to the lattice, governed by inelastic phonon scattering. This balance establishes the electron temperature that is higher than that of the lattice~\cite{Kravtsov2025}. The elevated electron temperature can be detected in resistance measurements, owing to the sensitivity of electron mobility, quantum oscillations, etc., to the electron temperature. In our experiments, resonant electron heating in the region of CR primarily suppresses the SdH oscillations since the value of $|B|=B_c$ corresponds to strong Landau quantization where the SdH effect is well developed. Reducing the SdH amplitude through heating leads to an oscillating photoresistance signal, $\Delta R$, with a phase that is opposite to that of the SdH oscillations in the non-illuminated $R_{xx}$, as shown in Fig.~\ref{fig1}c. The dashed line in Fig.~\ref{fig1}d, which corresponds to $B=B_c$ at different $n$, shows that the maximum amplitude of the oscillations in $\Delta R$ indeed corresponds to the expected CR position.

The central distinction between the CR and BM resonances is that excitation of the Bernstein modes is possible only in the presence of short-wavelength components of the electromagnetic field corresponding to the avoided crossings in the magnetoplasmon dispersion in Fig.~\ref{fig1}a. Under the conditions of our experiments, the wave-vector mismatch between the normally incident THz radiation and Bernstein magnetoplasmons is overcome by near fields generated via diffraction of the plane THz wave on the contacts that penetrate the graphene sample area. The near field that would occur in the vicinity of the contacts in the absence of graphene, has a broad distribution in $q$-space. Only those $q$-components of the field that are close to the magnetoplasmon dispersion in Fig.~\ref{fig1}a are strongly coupled to the electron motion in graphene and give rise to a resonant plasmon enhancement in the form of coupled density and electric field oscillations emanating from the contacts.\footnote{This is the main distinction from more traditional plasmonic experiments, where excitation of plasmons is provided by a periodic metal grating. There, the wave vector $q=2\pi/L$ of excited plasmons is fixed by the period $L$ of the grating.\cite{Maier2007}} Moreover, these oscillations become especially strong in the vicinity of the BM gaps at $\omega/\omega_c\simeq$~2, 3, $\ldots$,\cite{BandurinBernstein} featuring an almost flat dispersion of magnetoplasmons, and thus corresponding to an extremely high density of excited modes. The BM photoresponse, therefore, combines the near-field enhancement with the usual plasmonic amplification and additional resonant amplification due to flat dispersion in the vicinity of the BM gaps. This gives rise to a strong BM signal despite the relatively small area of the sample where the BM-mediated near field is present and, as we will discuss below, leads to bleaching of the BM resonances emerging at very low intensities of THz radiation. Similar to the CR, the electron scattering leads to resonant energy absorption from the excited BMs\cite{Dorozhkin2021,BandurinBernstein} and, via the energy balance in the stationary state, to a resonant heating of electrons that can be detected in the photoresistance measurements. However, unlike the CR where the heating is uniform, the BM resonances arise due to local electron heating in the regions near the contacts.

In  the region of CR, the electron heating in our experiments results primarily in the suppression of the SdH oscillations, since the value of $B_c$ corresponds to quantizing magnetic fields where the SdH effect is already well developed. The reduction of the SdH amplitude due to heating leads to an oscillating photoresistance signal, $\Delta R$, with a phase opposite to that of the SdH oscillations in the non-illuminated $R_{xx}$, see Fig.~\ref{fig1}c. The dashed line in Fig.~\ref{fig1}d, corresponding to $B=B_c$ at different $n$, shows that the maximal amplitude of oscillations in $\Delta R$ indeed corresponds to the expected CR position. %\footnote{It should be taken into account, however, that the exponential growth of the SdH oscillations can modify the shape of CR of $\Delta R$ in respect with that in absorption, making it broader and asymmetric.}  
On the other hand, the BM photoresponse, at low $\omega$ and $I$, appears as a single resonant peak. Since the SdH oscillations are exponentially suppressed at the corresponding weaker $B\sim B_c/2$, electron heating at such low $B$ primarily leads to a decrease in mobility, resulting in a resonant increase of resistivity and a positive peak in the photoresistance $\Delta R$, as illustrated in Fig.~\ref{fig1}c. %\footnote{We mention that this mechanism also provides a sub-leading contribution to the CR photoresponse which, however, is not important for our analysis and conclusions since the CR response is mostly determined by the heating-induced suppression of the SdH oscillations.} 
Quantum oscillations can still affect the shape of Bernstein resonances at higher $\omega$ and $I$, see Figs.~\ref{fig2}d, ~\ref{fig3}a and discussion below. On the other hand, the photoresponse in the vicinity of the CR contains sub-leading non-oscillatory components. These are not important for our analysis, which is based on the dominant oscillatory CR photoresponse.

We now turn to the observed nonlinear intensity dependences of the photoresistance and discuss the similarities and distinctions in the behavior of the CR and BM resonances. The saturation (bleaching) of the CR photoresponse produced by electron heating has been studied in many works on different materials and is well understood.\cite{Allan1985,Ganichev2005}  In the semiclassical regime, it is usually caused by a rapid increase of the inelastic relaxation rate with growing electron temperature. Effectively, this accelerates the rate of energy transfer to the lattice at high intensities, leading to a slower increase of the electron temperature in the nonlinear regime. This results in saturation of the intensity dependence of the photoresistance, which is frequently well described by Eq.~\eqref{formula}, see, e.g., Refs.~\cite{Allan1985,Ganichev2005}, and, for graphene, Ref.~\cite{Candussio2021a}.  

Our analysis in Figs.~\ref{fig2} and~\ref{fig3} shows that Eq.~\eqref{formula} also describes well the nonlinear behavior of BM resonances at overtones of the CR. The most remarkable finding is that the BM resonances, while producing a strong signal in $\Delta R$ at low $I$, of magnitude similar to $\Delta R$ at the CR, enter the nonlinear regime of strong saturation much earlier than the CR. Though surprising at first glance, this result in fact correlates well with the near-field nature of the BM signals and their strong enhancement in the vicinity of the BM gaps, where the spectrum of magnetoplasmons is almost flat. Due to the BM-mediated strong near-field enhancement, the short-wavelength THz field component in the vicinity of the contacts has a much higher amplitude than that of the incoming plane THz wave. On one hand, this yields the observed strong BM signal at low $I$, which is comparable in magnitude to the CR signal collected from the entire sample, since the strong near field of BMs leads to much stronger heating in the vicinity of contacts (hot spots) compared to the rest of the sample. On the other hand, stronger heating within hot spots naturally leads to saturation of the BM response in $\Delta R$ at much lower intensities of incoming radiation than the CR response. Therefore, the low saturation intensities of the BM resonances compared to CR provide strong evidence for the BM-mediated near-field nature of the photoresponse at the overtones of the CR. 

Proceeding to a more detailed discussion of the observations, Fig.~\ref{fig2}d shows that the shape of the BM signal in $\Delta R$ changes in the saturated regime, where one observes a progressively stronger admixture of magnetooscillations on top of the resonant peak at $B_c/2$. At first glance, this could be caused by the low-$B$ tail of the CR, which continues to grow almost linearly in the saturated regime of the BM response. On the other hand, the amplitude of magnetooscillations at the highest intensities is abruptly enhanced at $B_c/2$ and remains almost constant over a substantial region of $B$ above $B_c/2$. While the exact reason for that is currently unclear, the origin may be that the hot spots created by the near-field excitation of the BM magnetoplasmons expand at high intensities due to thermal diffusion, producing a suppression of the SdH oscillations over a larger area of the sample. This behavior, apparently related to the nonlinear interplay of spatially inhomogeneous temperature and BM field profiles, defines an interesting direction for future studies.

We continue with a discussion of the frequency dependence of the BM response, summarized in Fig.~\ref{fig3}d. It shows that the linear amplitude of the BM peaks at $B_c/2$, $\Delta R=A(f) I$, decreases with increasing $f$, while the saturation intensity $I_s(f)$ exhibits the opposite behavior. This is not surprising because, within the saturation mechanism leading to Eq.~\eqref{formula}, both the low-intensity linear coefficient $A(f)$ and the inverse saturation intensity $1/I_s(f)$  are proportional to the low-$I$ absorption cross-section. On the other hand, since the linear photoresistance signal $A(f) I$ is determined by the amount of energy absorbed via the near-field BMs, its frequency dependence should be governed by both, the absorption cross section and the effectiveness of the BM-assisted near field generation at the contacts, the latter being sensitive to the particular experimental configuration. Since these two origins of the frequency dependence cannot be disentangled, the functional behavior of $I_s(f)$ and $A(f)$ does not have fundamental significance.  It may change if a different setup is used for excitation of the Bernstein modes. From this perspective, our most important observation concerning the frequency dependence is that the saturation intensity $I_s(f)$ behaves approximately as the inverse of the linear coefficient $A(f)$. More systematic studies of the frequency dependence would be possible with help of tunable frequency radiation sources, e.g., a free electron laser. We also mention that the increase of measurement temperature should lead to growth of the saturation intensity, since it is proportional to the energy relaxation rate that is controlled by the phonon scattering and grows with temperature. This is the reason why in Ref.~\cite{BandurinBernstein} measurements of the intensity dependence on a similar sample at $T=20$~K did not shown any nonlinearities. In the Appendix \ref{appendixA1}, we include a new analysis of these data following the methodology developed in the present work. It shows that the intensity dependence at high temperature of $T=20$~K remains linear in both regions of BM and CR resonances, unlike the case of low $T=1.8$~K, Fig.~\ref{figS3}. 

Finally, we turn to a discussion of the polarization studies of the BM photosignals. In the case of circular polarization of the incoming THz radiation, Fig.~\ref{fig4}a, the photoresponse in the vicinity of the CR is strongly enhanced for the CR-active helicity (red curve), while the BM resonance has the same magnitude and shape for both helicities. This can be naturally attributed to the insensitivity of the diffracted high-$q$ near-field components of the THz field to the helicity of incoming radiation. Indeed, the BM magnetoplasmons are longitudinal waves with a (locally) well defined direction of propagation and thus have lower symmetry compared to the uniform CR mode, which has no such axial asymmetry. 

In general, the diffraction on contacts may also be sensitive to the direction of the linearly-polarized radiation, characterized by the angle $\alpha$ of the external THz electric field, while the CR is expected to be $\alpha$-independent. These expectations are clearly confirmed by the measurements presented in Figs.~\ref{fig4}b-e, including a strong $\cos^2 \alpha$ component, characteristic of the antenna effect, Figs.~\ref{fig4}c-e, which is observed in the $\alpha$-dependence of $\Delta R$ at $B\sim B_c/2$ for all three frequencies.

The data presented in Fig.~\ref{fig4}c-e show that the maximum signal is obtained when the electric field vector is oriented along the source-drain line, and that the overall polarization dependence is well described by the $\cos^2 \alpha$ function, which resembles the antenna effect. At the same time, we observe that the signal does not drop to zero at $\alpha=90^\circ$, as it would be expected for the antenna effect. Rather, it remains relatively large, at around 30\% of the maximum value. This is not surprising, as the pure $\cos^2 \alpha$ behavior is expected in the far field, while here we are dealing with the near field where the polarization dependence is more complex and not uniform, and antenna and diffraction effects cannot be clearly distinguished. An additional argument is that strong antenna effect in the far field would also lead to pronounced $\cos^2 \alpha$ modulation in the region of the CR, where we do not detect any $\alpha$-dependence of the photoresponse, see Fig.~\ref{fig4}(b). Moreover, the typical wave vectors of the Bernstein modes calculated in Ref.~\cite{BandurinBernstein} (about 1~$\mu$m$^{-1}$) indeed coincide with the inverse size of contacts overlaying graphene flakes. We also mention that sensitivity of the near field to the polarization angle may be either weak or strong depending on fine details of particular contacts and measurement geometry. For instance, in preliminary data obtained on a different sample (see inset in Fig.~\ref{figS4}) and presented in Supplemental Material to Ref.~\cite{BandurinBernstein} no significant difference of the BM response  at $\alpha=0$ and $90^\circ$ has been detected. More systematic studies of the linear polarization dependence with various configurations of contacts are desirable and should bring more detailed understanding of the near-field effects associated with excitation of Bernstein modes.

The observed essential differences in polarization behavior of the photoresistance at CR and its harmonics provide clear evidence for the near-field origin of the photoresponse at $B_c/2$ and $B_c/3$, and therefore, support their interpretation as a consequence of the resonant excitation of the BMs.

\section{Conclusion}
\label{Conclusion}
%{\bf Conclusion.}
In conclusion, we have observed strongly nonlinear THz photoresponse associated with the resonant excitation of Bernstein magnetoplasmons in graphene. We attribute the observed behavior to strong local heating of the electron gas in the vicinity of the metallic contacts in graphene, governed by a combination of  the field-concentration effect of the near field and plasmonic amplification. The near-field diffraction of the incoming plane-wave THz radiation on these contacts overcomes the wave-vector mismatch and enables the generation of short-wavelength magnetoplasmons. Moreover, the flat dispersion of the Bernstein modes provides significant enhancement of the near field in the region of gaps at the overtones of the cyclotron resonance. This BM-mediated enhancement leads to the formation of hot spots near the contacts and strong nonlinearities related to saturation of local electron heating. As a result, the saturation intensities for the BM resonances are much lower than those for the CR, governed by weaker uniform heating of the entire electron system. Further evidence of the near-field nature of the BM photoresponse is provided by polarization-sensitive studies. Our findings open a new direction for exploring nonlinear, non-local electrodynamics in 2D materials.

\section{Methods}
\label{Methods}
%\re{supplementary: add which contacts were used in particular measurements}

The measurements were performed applying monochromatic radiation of an optically pumped continuous wave ($cw$) molecular gas laser. The laser operated at frequencies $f = 0.69$~THz ($\lambda = 432$~$\mu$m and $\hbar\omega = 2.85$~meV), $f = 1.63$~THz ($\lambda = 184$~$\mu$m and $\hbar\omega  = 6.74$~meV) or $f = 2.54$~THz ($\lambda = 118$~$\mu$m and $\hbar\omega = 10.5$~meV). The  spot size of the Gaussian beam was monitored by a pyroelectric camera~\cite{Herrmann2016}. The diameter of the laser spot (1.5-3~mm) was much larger than the size of the graphene flake, ensuring uniform irradiation. In most of experiments, the radiation was linearly polarized. The initial state of polarization ($\alpha = 0, \varphi = 0$) pointed along the Hall bar, see the dashed line in the inset of Fig.~\ref{fig4}b. To analyze the polarization dependences of the photoresponse we used crystal quartz $\lambda/2$ or $\lambda/4$ wave plates. To vary the radiation power we used a cross-polarizer set-up, which consists of two wire grid polarizers. Rotation of the first one modified the radiation power, while the second one was fixed to ensure an unchanged output polarization~\cite{Hubmann2019,Candussio2021a}.

For photoresistance measurements, an alternating bias voltage ($U_\text{in} = 1$~V) at a  frequency $f_{ac} = 7.0$~Hz was applied across a load resistor ($R_L$ = 10~M$\Omega$) and the sample, see Fig.~\ref{fig1}b. Simultaneously,  THz radiation was modulated using an optical chopper operating at a frequency of $f_{op} = 215$~Hz. Such setup implements a double modulation technique, see Refs.~\cite{Kozlov2011,Otteneder2018} and Supplemental Material in Ref.~\cite{Savchenko2021}. It uses two lock-in amplifiers in series to yield a photoresistance signal, which changes its sign upon switching the bias voltage polarity. The first lock-in amplifier is phase-locked to the chopper frequency, isolating the photoresponse from the total signal. Its output contains two components: (i) a constant (DC) component, which is  proportional to the photocurrent and (ii) a component modulated at $f_{ac}$, which is proportional to the photoresistance.
This output serves as the input for the second lock-in amplifier, which is phase-locked to $f_{ac}$. It extracts the amplitude of the photoresistance signal, yielding a voltage denoted as $V$. The photoresistance is then extracted from the measured photosignal $V$ as $\Delta R=V R_L/\mid U_\text{in} \mid$.

%The radiation was modulated by an optical chopper at a frequency 215\,Hz and the photovoltage $U$ was measured using the standard lock-in technique. The signal was obtained by measuring the response to positive and negative bias and calculating the photoconductive signal as $U = (U_{\rm +3V} -  U_{\rm -3V})/2$ \re{Check values}. In doing so, we insure that the signal stems from the change of conductivity and not from any kinds of photocurrents. Indeed, while the photocurrent is independent of the polarity of the bias voltage, the photoconductive response, according to Ohm's law, changes its sign when the polarity of the bias voltage is reversed. In some measurements  we applied also  and 4-terminal configuration \re{(e.g. results presented in Fig.~\ref{fig1})} he double modulation technique, see Refs.~\onlinecite{Kozlov2011,Otteneder2018} and Supplemental Material in Ref.~\onlinecite{Savchenko2021}. 

The sample was mounted in a temperature-regulated Oxford Cryomag optical cryostat equipped with $z$-cut crystal quartz windows, which were covered by black polyethylene films that are almost transparent in the THz range, but block uncontrolled illumination from ambient light. Photoresistance  measurements were performed in Faraday geometry, with a magnetic field $B$ up to 4.0~T applied perpendicular to the graphene plane, see Fig.~\ref{figS1}. Results  presented in Fig.\ref{fig1}c were obtained in the 4-terminal geometry shown in Fig.~\ref{fig1}b, and those shown in Figs.~\ref{fig1}d and \ref{fig2}-\ref{fig4} in the 2-terminal geometry using the source and drain contacts to graphene.

\section{Acknowledgments}
\label{Acknowledgments}
The support of the Deutsche Forschungsgemeinschaft (DFG, German Research Foundation) via project WU 883/4-1  and by the European Union through the ERC-ADVANCED grant TERAPLASM No. 101053716 is gratefully acknowledged.  DFG is acknowledged for  support  via projects WU 883/3-1 (I.Y. and S.D.G.), DM~1/6-1 (I.A.D.), GU 2528/1-1 695298 (J.G.-S.), ID 314695032 - SFB 1277 subproject A09 (J. E. and M. M.) and ID 452301518 (J.W.). 
D.A.B. and A.S. acknowledge the support of the Singapore Ministry of Education Tier 2 grant award T2EP50123-0020 (photoresponse measurements) and AcRF Tier 1 grant 22-5390-P0001. 
W.K. acknowledges the support of “Center for Terahertz Research and Applications (CENTERA2)” project (FENG.02.01-IP.05-T004/23) carried out within the “International Research Agendas” program of the Foundation for Polish Science, co-financed by the European Union under European Funds for a Smart Economy Programme.
J.W. thanks funding from the European Union's Horizon 2020 research and innovation programme under the Marie Skłodowska-Curie grant agreement no. 861300. 
P.S., B.S., and I.G. thank Grant No. PAN.BFB.S.BWZ.369.022.2023 of the Polish Academy of Sciences and U.S. National Academy of Sciences.
%, awarded under the “Long-term program of support of the Ukrainian research teams at the Polish Academy of Sciences carried out in collaboration with the U.S. National Academy of Sciences with the financial support of external partners.”

\appendix
\counterwithin{figure}{section}
\setcounter{figure}{0}

\section{Sample design, transport characteristics, additional data on the density and intensity dependences of photoresistance}
\label{appendixA1}

\begin{figure*}[h]
	\centering
	\includegraphics[width=0.9\linewidth]{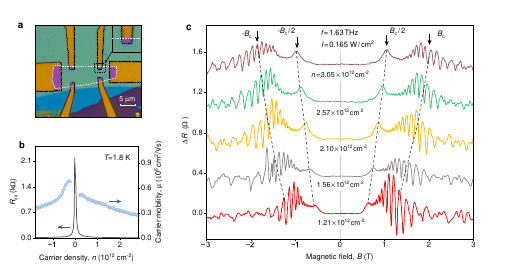}
	\caption{%\re{(2-terminal; measured on source-drain contacts)} (a) 
		Optical image of the sample. The dashed-line polygon indicates the edge of the graphene flake. A zoomed image in the top right corner show how contacts penetrate the sample area.  (b) Carrier density dependence of the $B=0$ longitudinal resistance $R_{xx}$ measured at  $T=1.8$ K (black line), combined with the correspondent carrier mobility $\mu$ (blue line). The carrier density was extracted from the Hall measurements of $R_{xy}$ at low $B$ (not shown). (c) Photoresistance $\Delta R$ versus the magnetic field, measured at $T=1.8$ K and $f=1.63$ THz for various carrier densities (an extended version of Fig.~1d in the main text). The densities $n$, as provided below each trace, were extracted from SdH oscillations period. The dashed lines indicate the position $B_c$ of CR and its main overtone $B_c/2$ at the corresponding densities. }
	\label{figS1}
\end{figure*}

Optical image of the sample is presented in Fig.~\ref{figS1}a. The exfoliated graphene flake was sandwiched between two hBN layers and equipped with Si$^{++}$ back gate giving access to a broad range of electron densities $n$. We used Si$^{++}$ substrate as a bottom gate with 285~nm SiO$_2$ and bottom hBN layer serving as a gate insulator. The high-quality hBN material was fabricated following the recipe in Ref.~\cite{Sadovyi2025}. On top of the structure, Cr/Au contacts were first defined by electron beam lithography and then the reactive ion etching by SF$_6$ gas was applied to the exposed area to remove the top hBN layer. Afterwards, 0.5~nm Cr and 20~nm Au were deposited in a physical vapor deposition chamber. Importantly, the side contacts penetrate the area of the graphene Hall bar, ensuring the near-field effects leading to excitation of the BM magnetoplasmons.

Figure~\ref{figS1}b shows the $B=0$ resistance of the sample and correspondent mobility as a function of the electron density $n$ varied via application of the gate voltage $V_g$. 
High mobility $\mu\lesssim 0.7 \times 10^6$ cm$^2$/V s of the sample confirms the high quality of the obtained structure. The connection $n = 0.68 \times10^{11}~V_g$~cm$^{-2}$V$^{-1}$ was established from the period of the SdH oscillations. Within the experimental accuracy, the same dependence $n(V_g)$ was determined from the Hall measurements at weak $B$ (not shown). We than used the standard value of the Fermi velocity from the literature, $v_F=10^6$~m/sec, to find the cyclotron frequency at the Fermi surface (and the corresponding cyclotron mass) at a given gate voltage. In this way we obtain dashed lines in all figures showing the position of the cyclotron resonance, $B_c=\hbar\omega \sqrt{\pi n}/(e v_F)$, and its harmonics. We underline that we do not obtain the electron mass from the cyclotron or BM resonances but rather use standard methods to determine their expected position, thus checking the consistency of our interpretation of results.

\begin{figure*}[h]
	\centering
	\includegraphics[width=\linewidth]{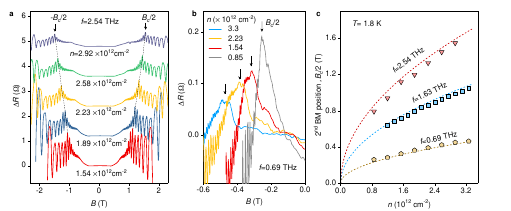}
	\caption{ %\re{(2-terminal; measured on source-drain contacts)} 
		(a) and (b)  Photoresistance $\Delta R$ versus the magnetic field $B$, measured at $T=1.8$ K for various carrier densities using $f=2.54$ THz (a) and $f=0.69$ THz (b) radiation. The densities $n$, as provided for each trace, were extracted from the SdH oscillations period. The arrows and dashed lines indicate the position $B_c/2$ at the corresponding densities. (c) Positions of the maxima in $\Delta R$ at the $B_c/2$ BM resonance extracted from measurements at various carrier densities for all three frequencies, as partially presented in Figs.~\ref{figS1}c and~\ref{figS2}a,b (symbols). The dashed lines show the calculated position of $B_c/2$ at these frequencies.}
	\label{figS2}
\end{figure*}

Figure~\ref{figS1}c shows the photoresistance traces for different densities, partially presented in Fig.~\ref{fig1}d of the main text, but now in the full range of magnetic fields. The photoresistance traces, obtained for two other radiation frequencies, are shown in Figs.~\ref{figS2}a and b. Figure~\ref{figS2}c collects the positions of the peaks corresponding to the primary BM resonance at $B=B_c/2$ for all three frequencies as function of the electron density $n$. Dashed lines plotted using the expression $B_c/2=\hbar\omega \sqrt{\pi n}/(2 e v_F)$ fit well the experimental data.

As an additional remark, we intentionally limited our study to the region of high electron densities where interaction effects on the electron spectrum are suppressed and the     standard value of the Fermi velocity can be used, in agreement with our findings. In future, sharp Bernstein resonances may indeed provide valuable information on the electron mass in the more interesting region of low densities near the neutrality point. However, it is a priori unclear how interactions should modify collective short wavelength excitations such as Bernstein modes (beyond the trivial quasistatic Coulomb repulsion). Apart from that, such studies require lower excitation frequencies to avoid the quantum Hall regime of transport, and must somehow avoid the strong inhomogeneity of the electron liquid related to the presence of contacts penetrating the sample area, which can be a nontrivial task in the region of ultra low densities. 

\begin{figure*}[h]
	\centering
	\includegraphics[width=0.5\linewidth]{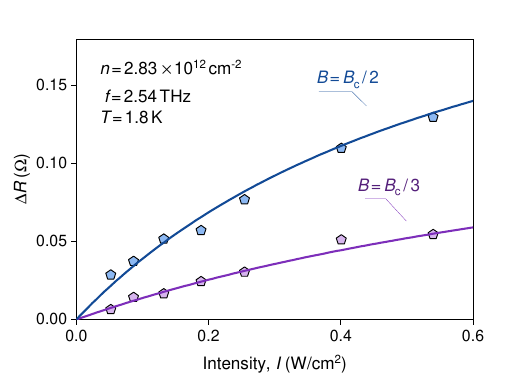}
	\caption{ Zoomed plots, which have been presented in Fig.~3b of the main text, show intensity dependences of the resonant photoresponse $B_c/3$ and $B_c/3$ excited by the radiation with $f=2.54$~THz. Solid lines are fits after Eq.~(1) of the main text.
		%nd  for $B_c/3$ and $B_c/3$ . %is represented by purple pentagons. Blue pentagons show the corresponding intensity evolution of resonances at $B_c/2$. Zoomed plots demonstrating saturation of the signals for $B_c/2$ and $B_c/3$ are presented in the Supplemental Materials
	} 
	\label{figS3}
\end{figure*}

Figure~\ref{figS3} shows zoomed plots in Fig.~\ref{fig3}b of the main text, demonstrating the sublinear growth (saturation) of the THz photoresponse for $B_c/2$ and $B_c/3$.

\begin{figure*}[h]
	\centering
	\includegraphics[width=0.95\linewidth]{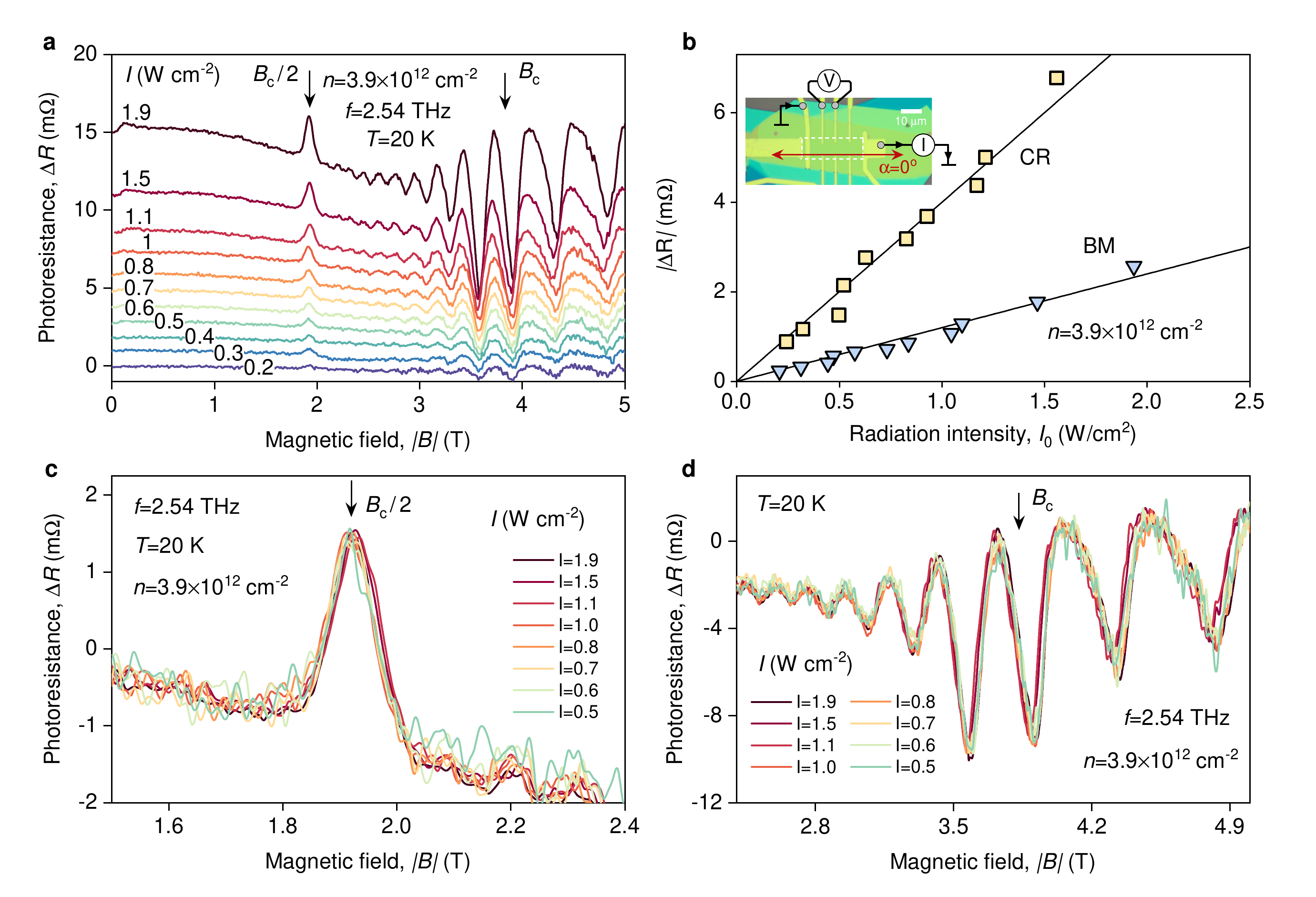}
	\caption{Intensity dependence on a similar sample at $T=20$~K and frequency $f=2.54$~THz presented in Fig.~S5 of the supplemental material to Ref.~\cite{BandurinBernstein}. (a) Photoresistance $\Delta R$ vs. magnetic field $B$ measured at given $n$ and $T$ and varied intensities $I$ of incoming irradiation with $f=2.54$ THz.  The corresponding intensity dependence of the photoresponse at CR ($B_c$) and BM ($B_c/2$). Triangles and squares are obtained from the rescaled traces near $B_c/2$ (panel c) and $B_c$ (panel d), respectively. Scaling factors are given in both panels.  Inset in panel (b) shows photograph of the sample and the experimental circuit used for measuring the photoresistivity. 
	} 
	\label{figS4}
\end{figure*}

Figure~\ref{figS4}a shows the intensity dependence of the photoresistance measured on a similar sample at $T=20$~K and frequency $f=2.54$~THz (Fig.~S5 of the supplemental material to Ref.~\cite{BandurinBernstein}). New analysis of these data in Fig.~\ref{figS4}b-d, using the same methodology as in Figs.~\ref{fig2} and \ref{fig3}, shows that, unlike the case of low $T=1.8$~K, Fig.~\ref{figS3}, here the photoresistance remains linear in the whole interval of intensities of incoming THz radiation, in both regions of BM and cyclotron resonances. We attribute this to a faster inelastic relaxation at $T=20$~K leading to a higher saturation intensity $I_s$. 

\section{Additional data on the polarization dependence}
	\label{appendixA2}
\begin{figure*}[h]
		\centering
		\includegraphics[width=0.8\linewidth]{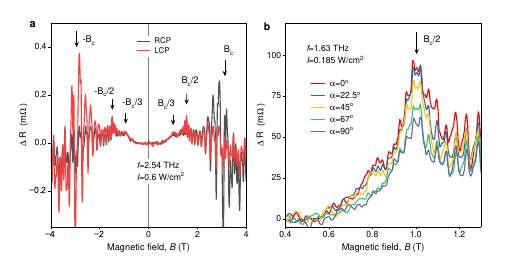}
		\caption{ (a) Photoresistance  $\Delta R$ vs the magnetic field $B$ in response to circularly polarized $f=2.54$~THz radiation of intensity $I=0.6$ W/cm$^2$, right-handed (RCP, gray line) and left-handed (LCP, red line). Arrows highlight the positions of the CR ($B_c$, highly sensitive to the helicity) and two BM resonances ($B_c/2$ and $B_c/3$, insensitive to helicity). (b) Photoresistance  $\Delta R$ vs. $B$ in response to linearly polarized $f=1.63$~THz radiation of intensity $I=0.185$ W/cm$^2$. Traces of different colors correspond to different in-plane orientation of the THz electric field in the incoming wave, marked by the azimuthal angle $\alpha$. Changes of the magnitude of the primary BM resonance at $B_c/2$, see arrow, follow $\cos^2 \alpha$ behavior characteristic for the antenna effect, as given in Fig.~4d in the main text. The carrier concentration $n=2.83\times 10^{12}$ cm$^{-2}$ in both panels.} 
		\label{figS5}
	\end{figure*}

	Figure~\ref{figS5}a shows the photoresistance traces for both helicities of $f=2.54$ THz circularly polarized radiation. Similar to the data presented in Fig.~\ref{fig4}a of the main text for the $f=1.63$ THz radiation, it shows that both BM resonances at $B_c/2$ and $B_c/3$ are insensitive to the radiation helicity, whereas the CR is strongly enhanced for the CR-active helicity ($B=B_c$ for the RCP, and $B=-B_c$ for the LCP). A larger signal is detected at the CR for the CR-active combination of radiation helicity and magnetic field polarity. In the Faraday geometry used in our experiments, the circularly polarized electromagnetic wave only produces the CR for one magnetic field polarity. Specifically, the electrons should perform circular motion in the same direction as the driving force of the THz wave; only then do the electric and magnetic forces continuously match. This is known as the CR-active (CRA) regime. Consequently, the CR should be absent for the opposite polarity of the magnetic field or opposite helicity of the EM wave, which is usually called the CR-inactive (CRI) regime. The difference in photoresistance magnitude shown in Fig.~\ref{fig4}a is therefore caused by strong CR radiation absorption in the CRA geometry. In contrast, the BM resonance has the same magnitude and shape for both helicities. This can naturally be attributed to the diffracted high-$q$ near-field components of the THz field being insensitive to the helicity of incoming radiation. Indeed, BM magnetoplasmons are longitudinal waves with a well-defined local direction of propagation and thus have lower symmetry than the uniform CR mode, which has no axial asymmetry.
    
    In Fig.~\ref{figS5}b, the variation of the $B_c/2$ BM response with the angle $\alpha$ of linearly polarized $f=1.63$ THz radiation is shown. Similar to the data in Fig.~4b in the main text, the maximal BM response is observed at $\alpha= 0$ and $180^\circ$, reflecting the $\cos^2 \alpha$ component characteristic for the antenna effect, see also Fig.~\ref{fig4}d of the main text. 

%doi/10.1063/5.0272610
%\bibliography{all_lib1.bib}
\bibliography{Bibliography.bib}
\end{document}